\documentclass[preprint,superscriptaddress,amsmath,amssymb,prd,aps,showpacs,floatfix,nofootinbib]{revtex4-1}

\usepackage{graphicx}% Include figure filesa
\usepackage{dcolumn}% Aligntable columns on decimal point
\usepackage{bm}
\usepackage{mathrsfs}% bold math
\usepackage{hyperref}% add hypertext capabilities
\usepackage{amsmath}
\usepackage{amssymb}
\usepackage{bm}
\usepackage{color}
\usepackage{float}
\usepackage{dcolumn}
\usepackage{multirow}
\usepackage{changepage}
\usepackage{enumerate}
\hypersetup{pdftex,colorlinks=true,linkcolor=blue,citecolor=red,menucolor=black,urlcolor=blue,filecolor=blue}

\newcommand{\mev}{\textrm{ MeV}}

\begin{document}
%\title{$\boldsymbol{\Xi_{bb}}$ and $\boldsymbol{\Omega_{bbb}}$ molecular states}
%\title{Prediction of new $T_{cc}$ states of $D^{*} D^{*}$ and $D^{*}_s D^{*}$ molecular nature}
\title{Looking for the exotic $X_0(2866)$  and its $J^P=1^+$ partner in the $\bar{B}^0 \to D^{(*)+}  K^- K^{(*)0}   $  reactions  }
\date{\today}

\author{L.~R. Dai}
\email{dailianrong@zjhu.edu.cn}
\affiliation{School of Science, Huzhou University, Huzhou 313000, Zhejiang, China}
\affiliation{Departamento de F\'{\i}sica Te\'orica and IFIC, Centro Mixto Universidad de Valencia-CSIC Institutos de Investigaci\'on de Paterna, Aptdo.22085, 46071 Valencia, Spain}

\author{R.~Molina}
\email{Raquel.Molina@ific.uv.es}
\affiliation{Departamento de F\'{\i}sica Te\'orica and IFIC, Centro Mixto Universidad de
Valencia-CSIC Institutos de Investigaci\'on de Paterna, Aptdo.22085,
46071 Valencia, Spain}

\author{E.~Oset}
\email{oset@ific.uv.es}
\affiliation{Departamento de F\'{\i}sica Te\'orica and IFIC, Centro Mixto Universidad de
Valencia-CSIC Institutos de Investigaci\'on de Paterna, Aptdo.22085,
46071 Valencia, Spain}

\begin{abstract}
We propose two reactions,  $\bar{B}^0 \to K^0  D^{+}  K^-$ and  $\bar{B}^0 \to K^{*0}  D^{*+}  K^-$, already measured at Belle, to look into
the $J^P=0^+$, $X_0(2866)$ state and a   $1^+$  partner of molecular $D^*\bar{K}^*$  nature, by looking at the $D^+ K^-$ and  $D^{*+} K^-$
invariant mass distributions, respectively. Very clear peaks over the background are predicted and the  branching ratios for the production of these
states are evaluated to facilitate the task of determining the needed statistics for their observation. The fact that  the $K^{*0} K^-$ mass
distribution from accumulation of events of different reactions  is already available, indicates that one is not far away from having the needed
statistics to see peaks in the  $D^{(*)+}  K^-$ mass distributions, so far  not yet analyzed.
\end{abstract}

\maketitle

\section{Introduction}
%\cite{}
In the paper \cite{belle} the Belle collaboration reported on the $\bar{B}^0 \to D^{(*)+}  K^- K^{(*)0}$  decays, giving a list of eight reactions
for which the  branching ratios were provided. In some of the reactions:
a) $\bar{B}^0 \to D^{+}  K^- K^{*0}$,
b) $\bar{B}^0 \to D^{*+}  K^- K^{*0}$,
c) $\bar{B}^0 \to D^{+}  K^- K^{0}$,
d) $\bar{B}^0 \to D^{*+}  K^- K^{0}$,
one finds pairs,  $ D^+ K^-$ in $a)$ and $c)$,  $D^{*+} K^-$ in $b)$ and $d)$,
which contain open charm and strangeness  with $c$ and $s$ quarks.  Should these pairs result from the decay of a physical state, it would be
genuinely exotic since  it cannot come from a $q\bar{q}$ conventional meson.  The chosen pairs could correspond to states  with isospin $I=0$, while
the other four cases of \cite{belle} would correspond to $D^{(*)+} \bar{K}$ states  with isospin $I=1$. The limited statistics prevented to get  $D^{(*)+}  K^-$ mass distributions,
while the accumulation of $K^- K^{*0}$ events from four reactions allowed to get a  $K^- K^{*0}$ mass distribution that evidenced the $B \to D^{(*)+} a_1(1260)$
decay with $a_1(1260) \to K^- K^{*0}$. Yet, the abundant literature on tetraquark states from the very beginning of
the quark model \cite{rm1,rm2,rm3,rm4,rm5,rm8,rm9,rm7,rm6,rm11,rm10} up to now (see Refs. \cite{hxchen,ahmed,pilloni,rosner,xliu,nora} for reviews on more
recent works), would have made advisable to look at the  $D^{(*)+}  \bar{K}$ mass distributions in search of possible peaks corresponding to exotic states.

In between, the answer to this question was provided by the LHCb collaboration \cite{lhcb1,lhcb2} with the finding of the
 $X_0(2866)$ and $X_1(2900)$ states in the $B^+ \to D^+  D^- K^+$  decay by looking at the $D^- K^+$  invariant  mass distribution.
 In the charge conjugate reaction $B^- \to D^+  D^- K^-$ one would find the peaks  in the $D^+ K^-$  invariant  mass distribution.
 Interestingly, the existence of a   $I=0, J^P=0^+$ molecular state of $D^* \bar{K}^*$ nature, decaying to $D\bar{K}$,
  had been predicted in \cite{tania}, with a mass
 of $2848\mev$ and a width between $23-59\mev$, very close to the data of the $X_0(2866)$ with mass $2866\pm 7\mev$ and width $57.2 \pm 12.9\mev$. An update
 of that work to the light of the LHCb results is presented in  \cite{raquelx}.

 The findings of Refs.  \cite{lhcb1,lhcb2} prompted many works offering an explanation for the  $X_0(2866)$ as a tetraquark state \cite{d5,d6,d7,d8} or
 a  molecular $D^* \bar{K}^*$ state  \cite{d14,d15,d16,d17,d18,d19}. The sum rules studies \cite{d9,d10,d11,d12,d13} have also contributed its share to
 the discussion, some of them proposing a molecular  structure \cite{d11,d12,d13}. Other studies suggest a structure coming from a  triangle singularity \cite{d21} or cusps and analytical properties of triangle diagrams  \cite{d22,d23}. A triangle mechanism is also suggested  in \cite{qifang}, and in \cite{d20}
 a  detailed quark  model calculation is shown to disfavor the compact tetraquark picture.

 In  Ref. \cite{tania}, apart from the $J^P=0^+$ state, two other states with $J^P=1^+, 2^+$ also in $I=0$ were found. In the update of \cite{raquelx},
 where the free parameters of the model were adjusted to experiment \cite{lhcb1} for the  $X_0(2866)$, the masses, widths and couplings to the  $D^* \bar{K}^*$
 channel were evaluated, which are shown in  Table \ref{tab:1}.
\begin{table}[t]
\renewcommand{\arraystretch}{1.2}
\begin{center}
\caption{Properties of the $D^*\bar{K}^*$ states from  Ref. \cite{raquelx} accounting for $D \bar{K}$ and  $D^* \bar{K}$ decays.}
 \begin{tabular}{cccccc}
 \hline
 ~~~ $I [J^P]$ ~~~&~~~ $M~[\mathrm{MeV}]$ ~~~ & ~~~$\Gamma~[\mathrm{MeV}]$ ~~~ &~~~ Coupled channels~~~ &~~~ $g_{R,D^*\bar{K}^*}$ $[\mathrm{MeV}]$ &~~~ state~~~\\
  \hline
  $0 [2^+]$& $2775$ & $38$ & $D^*\bar{K}^*$ & $16536$ &?\\
  $0 [1^+]$& $2861$ & $20$ & $D^*\bar{K}^*$ & $12056$ & ?\\
  $0 [0^+]$&$2866$ & $57$& $D^*\bar{K}^*$ & $11276$ & $X_0(2866)$\\
  \hline
 \end{tabular}
\end{center}
\label{tab:1}
\end{table}

For reasons of parity and angular momentum conservations, the $0^+$ state only decays to $D \bar{K}$ while the $1^+$ state decays to $D^* \bar{K}$.

The purpose of the present work is to investigate  whether by looking at the $\bar{B}^0 \to D^{(*)+}  K^- K^{(*)0}$ reactions one can observe clear peaks in the
$D^{(*)+}  K^-$ spectrum. The reaction is similar to the $B^- \to D^+  D^- K^-$   one studied in \cite{lhcb1,lhcb2}. The $K^{(*)0}$ in the Belle
 reactions would play the role of the $D^-$ in the LHCb one.  The study is stimulated by the success found in \cite{newpaper}, fairly reproducing the
$D \bar{K}$  peak versus the background of \cite{lhcb1} in the study of the $B^- \to D^+  D^- K^-$  reaction. This success was used in \cite{newpaper}
to suggest the $\bar{B}^0 \to D^{*+}  D^{*0} K^- $ decay in order to investigate the  $1^+$  state of Table \ref{tab:1} by looking at the
 $D^{*+}  K^-$ mass distribution.  It was found that the  $1^+$  state generated a peak in the distribution with a strength about $7$ times bigger than the
 background at the peak of the $1^+$ contribution. Based on these findings we propose here to study the  $\bar{B}^0 \to D^{*+}  K^- K^{*0}$ and
  $\bar{B}^0 \to D^{+}  K^- K^{0}$ reactions. The reason to choose these two reactions between the eight reactions of Belle \cite{belle}  is that,
  both in the signal for the exotic states as in the background, the amplitudes can proceed in $s$-wave, assumed dominant as usual, and one
  can correlate the background and the signal for the production of the $0^+$ and  $1^+$  states.

\section{FORMALISM and  results}

\subsection{Production of the $1^+$  state in $\bar{B}^0 \to D^{*+}  K^{*-} K^{*0} \to D^{*+}  K^- K^{*0}$}
\begin{figure}[h]
\centering
\includegraphics[scale=0.85]{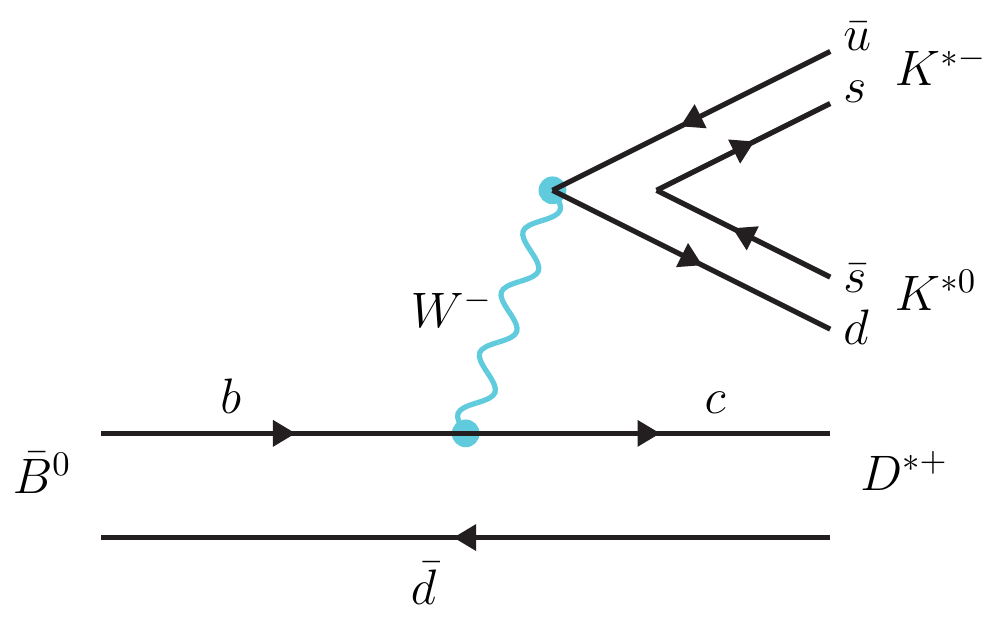}
\caption{Diagrammatic decay of  $\bar{B}^0 \to   D^{*+} K^{*-} K^{*0} $ at the quark level.}
\label{fig:1}
\end{figure}

\begin{figure}[h]
\centering
\includegraphics[scale=0.8]{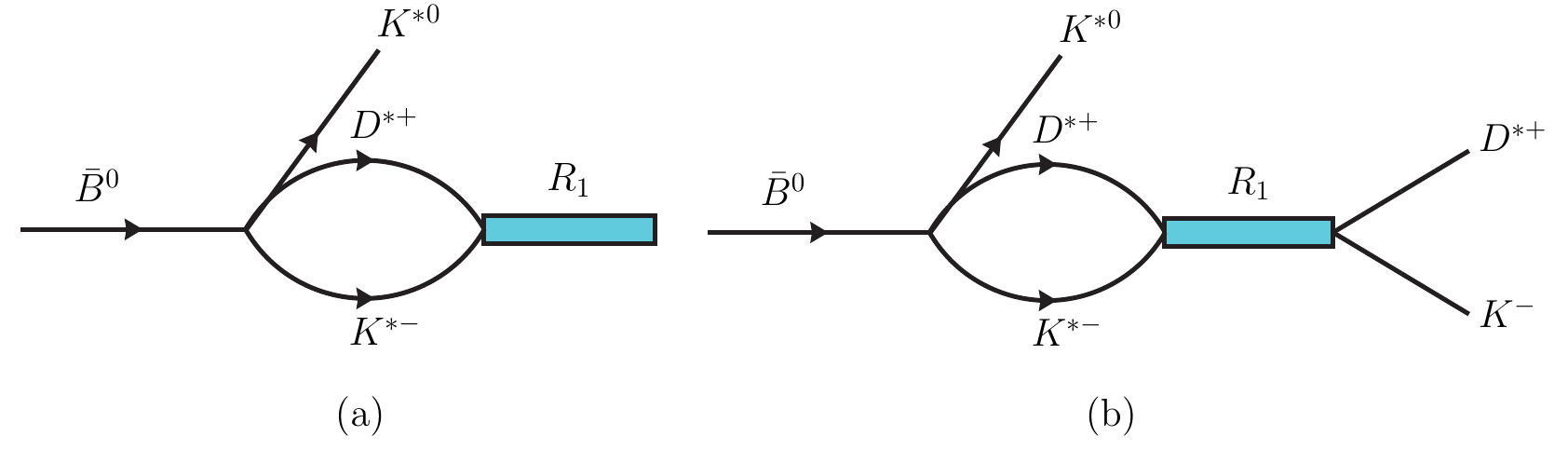}
\caption{(a) Rescattering of $D^{*+}  K^{*-}$ to give the resonance $R_1$;~(b) Further decay of $R_1$ into $D^{*+} K^-$.}
\label{fig:2}
\end{figure}

The $D^*\bar{K}^*$  $1^+$ state can only decay in $D^*\bar{K}$. Thus, we choose the  $\bar{B}^0 \to D^{*+}  K^- K^{*0}$ reaction and look at
$D^{*+}  K^-$ mass distribution. The signal, however, will come  from the $\bar{B}^0 \to D^{*+} K^{*-} K^{*0}$ reaction, after the final state
interaction of  $D^{*+} K^{*-}$ to give the  $1^+$ state ($R_1$) and posterior decay into $D^{*+}  K^-$. The primary step proceeds via external emission
 as depicted in Fig.~\ref{fig:1}.
The $\bar{u}d$ component after the $W^-$ vertex is hadronized with an $s\bar{s}$  component to give rise to $K^{*-} K^{*0}$ and the
$c\bar{d}$ gives the $D^{*+}$. One has three vectors and one can write the $s$-wave  component of the transition matrix
matching the angular momentum of the $\bar{B}^0$ as:
\begin{eqnarray}
 \tilde{t} = {\cal{C}}\, {\bm \epsilon^{(1)}}\cdot ({\bm \epsilon^{(2)}} \times {\bm \epsilon^{(3)}})= {\cal{C}}\, \epsilon_{ijk}\epsilon^{(1)}_i
\epsilon^{(2)}_j  \epsilon^{(3)}_k
\label{eq:t}
\end{eqnarray}
where the indices $1,2,3$ apply to the $K^{*0}$, $D^{*0}$ and $K^{*-}$, respectively. We observe how the spins of the particles $2$ and $3$  combine to $J=1$.
The next step is to consider the $D^{*+} K^{*-}$ interaction. With our phase convention
$(D^{*+},-D^{*0})$, $(\bar{D}^{*0},D^{*-})$,  $(K^{*+},K^{*0})$,  $(\bar{K}^{*0},-K^{*-})$, the $I=0$  $D^* \bar{K}^*$ state is written as
\begin{eqnarray}
|D^* \bar{K}^*; I=0\rangle = -\frac{1}{\sqrt{2}}(D^{*+}K^{*-}+D^{*0}\bar{K}^{*0} ) \,.
\end{eqnarray}
The final state interaction of $D^{*+} K^{*-}$ to produce the $R_1$ state is taken into account as shown diagrammatically in Fig.~\ref{fig:2}.

We also need the vertex $R_1 D^* \bar{K}^*$ which incorporates the spin projection generator $D^i$
\begin{eqnarray}
\tilde{g}_i=g_i V^{(i)}
\label{eq:g}
\end{eqnarray}
with $V^{(i)}$ given by \cite{liangmolina}
\begin{eqnarray}
V^{(0)}&=& \frac{1}{3} \epsilon^{(2)}_l \epsilon^{(3)}_l \delta_{ij} \nonumber \\
V^{(1)}&=& \frac{1}{2} \big(\epsilon^{(2)}_i \epsilon^{(3)}_j -\epsilon^{(2)}_j \epsilon^{(3)}_i \big)     \nonumber  \\
V^{(2)}&=&   \frac{1}{2} \big(\epsilon^{(2)}_i \epsilon^{(3)}_j +\epsilon^{(2)}_j \epsilon^{(3)}_i \big)- \frac{1}{3} \epsilon^{(2)}_l \epsilon^{(3)}_l \delta_{ij}
\end{eqnarray}
Considering the $\tilde{t}$ matrix of Eq.~\eqref{eq:t} and Eq.~\eqref{eq:g} for $V^{(1)}$,  $V^{(1)}=\frac{1}{2} \big(\epsilon^{(2)}_{i'} \epsilon^{(3)}_{j'} -\epsilon^{(2)}_{j'} \epsilon^{(3)}_{i'} \big)$, and that in the loop function we sum over the spin polarization
$\sum_{pol} \epsilon^{(r)}_i \epsilon^{(r)}_j =\delta_{ij}~(r=2,3)$, we obtain
\begin{eqnarray}
t = {\cal{C}}\, \epsilon^{(1)} \epsilon_{i i' j'} G_{D^*\bar{K}^*} (M_{\rm inv}(D^*\bar{K}^*))  \frac{-1}{\sqrt{2}} g_{R,D^*\bar{K}^*} \nonumber
\end{eqnarray}
where $g_{R_1,D^*\bar{K}^*}$  is the coupling of the resonance $R_1$ to the $(I=0)$ $D^*\bar{K}^*$  state
and $G_{D^*\bar{K}^*}$ is the loop function of the $D^*,\bar{K}^*$ integrating the propagators of the two particles. We use dimensional regularization
for this loop with $\alpha=-1.474$ for a chosen $\mu=1500\mev$ as was needed in \cite{raquelx} to obtain the right mass of the $X_0(2866)$  state.

The sum over the final vector polarization in $\sum |t|^2$ is implemented by summing  $\sum |t|^2$ over the indices $i', j'$ for the
implicit $VV$ components of $R_1$ and over the index $i$ to sum over the  $K^{*0}$ polarization. We find
\begin{eqnarray}
\sum_{pol}|t|^2 &=& 3\, {\cal{C}}^2 \, |g_{R_1,D^*\bar{K}^*}|^2 \, |G_{D^*\bar{K}^*} (M_{\rm inv} (D^*\bar{K}^*)) |^2   \nonumber
\end{eqnarray}

The next step is to consider the decay of $R_1$ into $D^{*+}  K^-$, as depicted in Fig.~\ref{fig:2} (b). This leads to a  $t'$ matrix
containing the coupling of $R_1$ to $D^{*+}  K^-$. This is accomplished by an effective coupling $g_{R_1,D^* \bar{K}}$ to the $D^* \bar{K}^*, I=0$ state,
such that the coupling to $D^{*+}  K^-$ is $\frac{-1}{\sqrt{2}} g_{R_1,D^*\bar{K}}$. To get the $g_{R_1,D^*\bar{K}}$ coupling we use the $R_1$ decay  width via
\begin{eqnarray}
\Gamma_{R_1}=\frac{1}{8\pi} \frac{1}{M^2_{R_1}} |g_{R_1,D^* \bar{K}}|^2 q_{\bar{K}}\,;\qquad q_{\bar{K}}=\frac{\lambda^{1/2}(M^2_{R_1},m^2_{D^*},m^2_{\bar{K}})}{2 M_{R_1}}\ ,
\end{eqnarray}
taking the value of $\Gamma_{R_1}$ from Table \ref{tab:1}. Hence
\begin{eqnarray}
\sum_{pol}|t'|^2 &=&\frac{6}{4} {\cal{C}}^2 |g_{R_1,D^*\bar{K}^*}|^2 |G_{D^*\bar{K}^*} (M_{\rm inv})|^2   \nonumber\\
&\times&|g_{R_1,D^* \bar{K}}|^2 |\frac{1}{M^2_{\rm inv} (R_1)-M^2_{R_1}+i M_{R_1} \Gamma_{R_1}}|^2
\end{eqnarray}
The  invariant mass distribution is then  given by
\begin{eqnarray}
\frac{d\Gamma}{dM_{\rm inv} (D^{*+}K^-)}=\frac{1}{(2\pi)^3} \frac{1}{4M^2_{\bar{B}^0}} p_{\bar{K}^{*0}} \tilde{p}_{K^-} \sum |t'|^2
\label{eq:t2}
\end{eqnarray}
where
\begin{eqnarray}
 p_{\bar{K}^{*0}}=\frac{\lambda^{1/2}(M^2_{\bar{B}^0},m^2_{\bar{K}^{*0}},M^2_{\rm inv}(D^{*+} K^-)}{2 M_{\bar{B}^0}}\,, \quad
  \tilde{p}_{K^-} =\frac{\lambda^{1/2}(M^2_{\rm inv}(D^{*+} K^-),m^2_{D^*},m^2_{\bar{K}})}{2 M_{\rm inv}(D^{*+} K^-)}
\end{eqnarray}

We would like to compare this mass distribution with the one of the background for the same reaction, $\bar{B}^0 \to  K^{*0} D^{*+} K^{-}$.
The process proceeds with the same topology as in  Fig.~\ref{fig:1}, changing   $K^{*-}$ by $K^-$.  As shown in \cite{vectorpseudo} the
difference between pseudoscalar and vector production can be taken into account by means of Racah coefficients of the same order of magnitude,
so approximately we can put for the $\bar{B}^0 \to  K^{*0} D^{*+} K^{-}$ background
\begin{eqnarray}
t={\cal{C}} \, {\bm \epsilon (K^*)} \, {\bm \epsilon (D^*)}  \nonumber
\end{eqnarray}
with the same constant ${\cal{C}}$ as in Eq.~\eqref{eq:t}, such that now the  background distribution is given by
\begin{eqnarray}
\frac{d\Gamma_{\rm bac}}{dM_{\rm inv} (D^{*+}K^-)}=\frac{1}{(2\pi)^3} \frac{1}{4M^2_{\bar{B}^0}} p_{K^{*0}} \tilde{p}_{\bar{K}} \,3\, {\cal{C}}^2
\end{eqnarray}

The assumption of taking the same constant ${\cal{C}}$ is supported by the results of \cite{newpaper}, reproducing fairly well the signal versus
the background of the LHCb experiment \cite{lhcb1}.

\begin{figure}[h]
\centering
\includegraphics[scale=0.75]{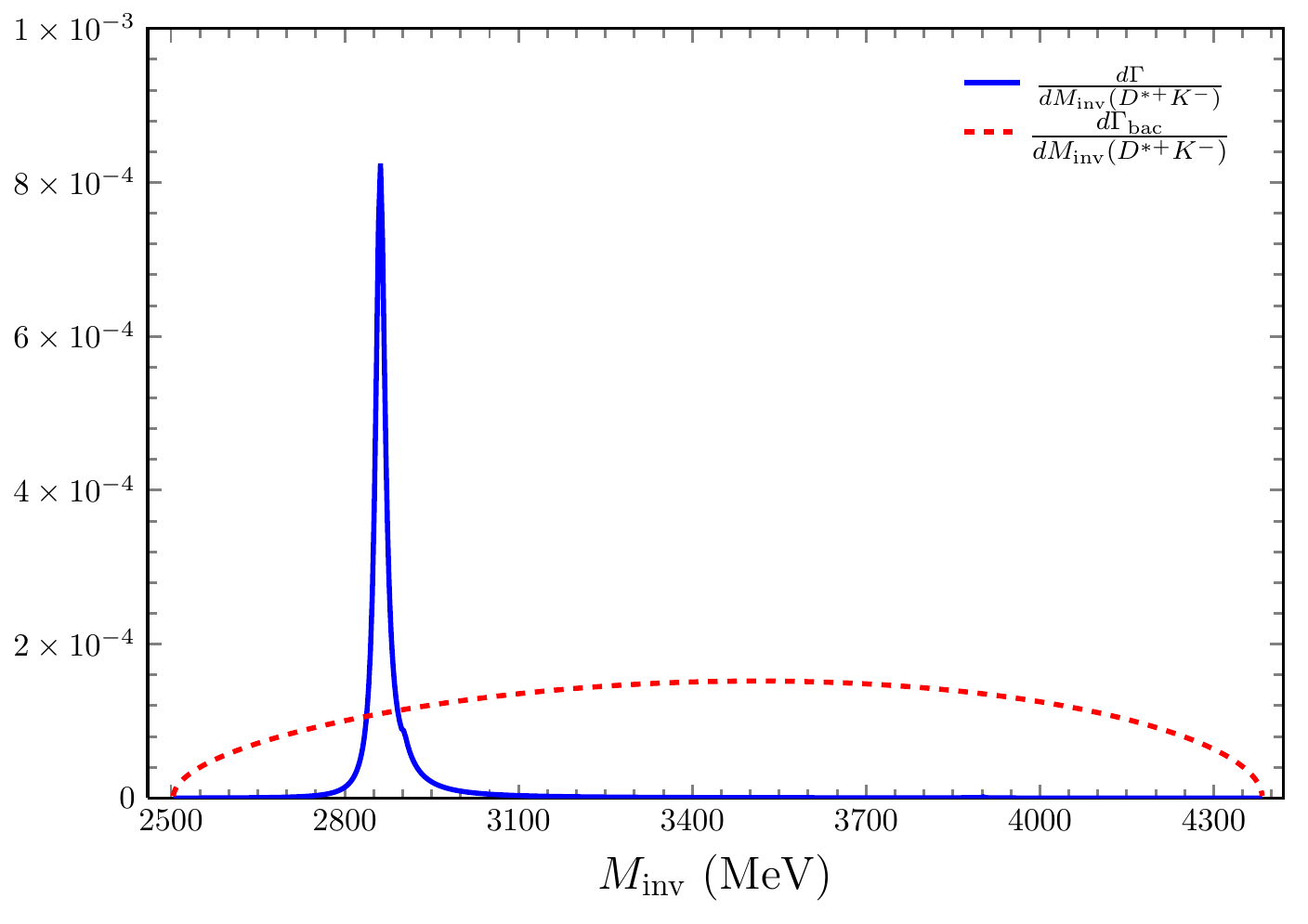}
\caption{$\frac{d\Gamma}{dM_{\rm inv}}$  for  $R_1$ production and  $\frac{d\Gamma_{\rm bac}}{dM_{\rm inv}}$  for background in the $\bar{B}^0 \to K^{*0} D^{*+} K^-$ reaction in global arbitrary units. $M_{\rm inv}$ is the $D^{*+} K^-$ invariant mass. }
\label{fig:dgR1}
\end{figure}

The results can be seen in Fig.~\ref{fig:dgR1}. We can see a peak clearly sticking out of the background, as was also  found in \cite{newpaper} with
a different reaction. It is clear that even if there were uncertainties of a factor of two or three, the signal should be clearly seen.

\subsection{Production of the $0^+$  state in $\bar{B}^0\to  K^0 D^{*+} K^{*-}\to K^0D^+K^-$}

Proceeding like in the former subsection, we would now compare the signal for the $0^+$  state from the
$\bar{B}^0 \to  K^0 D^{*+} K^{*-}$ with $D^{*+} K^{*-}$ interaction leading to the  $0^+$  state ($R_0$) and its decay
into $D^+ K^-$, and the background from the $\bar{B}^0 \to K^0  D^+ K^-$ reaction.  We can proceed as before and for the
$\bar{B}^0 \to K^0  D^{*+} K^{*-}$ we assume a transition matrix
\begin{eqnarray}
t={\cal{C'}} \, {\bm \epsilon (D^*)} \, {\bm \epsilon (K^*)}  \nonumber
\end{eqnarray}
and similarly for the $\bar{B}^0 \to K^0  D^+ K^-$
\begin{eqnarray}
t={\cal{C'}}  \nonumber
\end{eqnarray}
with the same ${\cal{C'}}$ as we have done before.  We shall come back to this assumption.  Following the same steps as before we obtain:
\begin{eqnarray}
\frac{d\Gamma'}{dM_{\rm inv} (D^{+}K^-)}=\frac{1}{(2\pi)^3} \frac{1}{4M^2_{\bar{B}^0}} p_{K^0} \tilde{p}_{K^-} \sum|t'|^2
\end{eqnarray}
where
\begin{eqnarray}
\sum|t'|^2 &=&\frac{3}{4} {\cal{C'}}^2 |G_{D^* \bar{K}^*} (M_{\rm inv}(D^+K^-)) |^2  \, |g_{R_0,D^* \bar{K}^*}|^2  \nonumber\\
&\times&  |\frac{1}{M^2_{\rm inv} (D^+ K^-)-M^2_{R_0}+i M_{R_0} \Gamma_{R_0}}|^2 \, |g_{R_0,D \bar{K}}|^2
\end{eqnarray}
with
\begin{eqnarray}
 p_{K^0}=\frac{\lambda^{1/2}(M^2_{\bar{B}^0},m^2_{K^0},M^2_{\rm inv}(D^{+} K^-)}{2 M_{\bar{B}^0}} \,, \quad
  \tilde{p}_{K^-} =\frac{\lambda^{1/2}(M^2_{\rm inv}(D^{+} K^-),m^2_{D^+},m^2_{K^-})}{2 M_{\rm inv}(D^{+} K^-)}
\end{eqnarray}
with  $g_{R_0,D^* \bar{K}^*}$ given in Table \ref{tab:1}, and the effective $|g_{R_0, D \bar{K}}|^2 $ coupling obtained from
\begin{eqnarray}
\Gamma_{R_0}=\frac{1}{8\pi} \frac{1}{M^2_{R_0}} |g_{R_0,D \bar{K}}|^2 q_{\bar{K}}\,;\qquad q_{\bar{K}}=\frac{\lambda^{1/2}(M^2_{R_0},m^2_{D^+},m^2_{\bar{K}})}{2 M_{R_0}}
\end{eqnarray}
For the background we find
\begin{eqnarray}
\frac{d\Gamma'_{\rm bac}}{dM_{\rm inv} (D^{+}K^-)}=\frac{1}{(2\pi)^3} \frac{1}{4M^2_{\bar{B}^0}} p_{K^0} \tilde{p}_{K^-} \, {\cal{C'}}^2
\end{eqnarray}

\begin{figure}[h]
\centering
\includegraphics[scale=0.75]{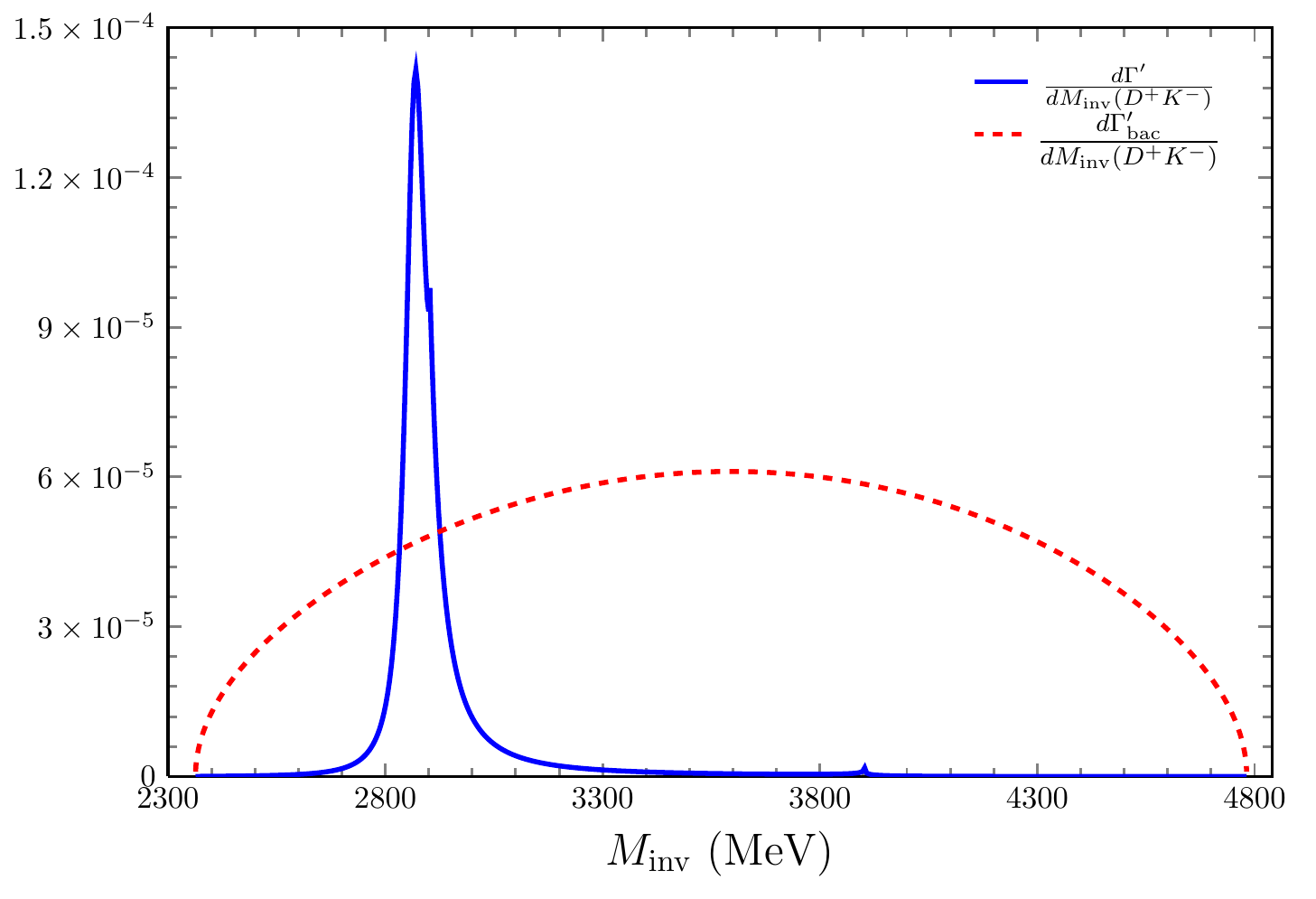}
\caption{$\frac{d\Gamma'}{dM_{\rm inv}}$  for  $R_0$ production and  $\frac{d\Gamma'_{\rm bac}}{dM_{\rm inv}}$  for background in the $\bar{B}^0 \to K^{+} D^{+} K^-$ reaction in global arbitrary units. $M_{\rm inv}$ is the $D^{+} K^-$ invariant mass. }
\label{fig:dgR0}
\end{figure}

The results for these two distributions are shown in Fig.~\ref{fig:dgR0}.  We also  see a signal that sticks out of the
background clearly, as was shown in \cite{newpaper} for the $B^- \to D^+  D^- K^-$  reaction in the production of the $X_0(2866)$.

As this point we would like to make some discussion. The $\bar{B}^0 \to K^0 D^{*+} K^{*-}$ can proceed with the topology of  Fig.~\ref{fig:1} changing $K^{*0}$ by $K^0$ where
the  $K^0$ and $K^{*-}$ are produced by hadronization of the $\bar{u}d$ component. Yet, the production of $K^0 K^-$ from the same vertex is suppressed,
as discussed in \cite{notwops}.  Indeed, the vertex $WPP$ is given by $W_\mu \langle [P,\partial_\mu P]\rangle$ in chiral theory \cite{chiral1,chiral2},
the $s$-wave going as the difference in the energies of the two pseudoscalars for the $WPP$ vertex, which  vanishes in the $W$ rest frame
if the particles have the same mass. The argument does not hold if one produces a vector and a pseudoscalar, as it was the case for the
signal of the $0^+$  state. The argument given above is corroborated by the branching ratio of the $\bar{B}^0 \to D^+  K^- K^0$, which is about
one order of magnitude  smaller than the one of $\bar{B}^0 \to D^{*+}  K^- K^{*0}$ (see table 2 of Ref. \cite{belle}).
Certainly we could now have contributions from higher partial waves, but the argumentation given above, with the support of the small
$\bar{B}^0 \to D^+ K^- K^0$  branching ratio, would tell us
that we can expect in practice a peak showing even stronger with respect to the background than what is shown in Fig.~\ref{fig:dgR0}.

In the LHCb case \cite{lhcb1} a different reaction was used, the $B^+ \to D^+  D^- K^+$, or analogously $B^- \to D^-  D^+ K^-$. Even
if the reaction seems  the same except for small changes as the $\bar{B}^0  \to K^0  D^+  K^-$, replacing the $D^-$  by $K^0$, the reactions
are topologically  different since the LHCb one, as well as the associated $B^- \to D^- D^{*+} K^{*-} $  reaction, proceed via internal emission
and the argument discussed above is peculiar of the $W^\mu \langle [P,\partial_\mu P]\rangle$ vertex of external emission. In the LHCb reaction  the
formalism used here  for the signal and background gave rise to a distribution in fair agreement with experiment. There is no analog reaction to the
$B^- \to D^-  D^+ K^-$ that proceeds with internal emission of the type $B \to D K \bar{K}$. The reaction that we have chosen to observe the
$0^+$, $X_0(2866)$ state, $\bar{B}^0 \to K^0 D^+  K^- $, stands as a good one, where the signal over background is expected to be even bigger than
shown in Fig.~\ref{fig:dgR0}.

% below old

\section{Conclusions}

We have chosen two reactions, already performed by the Belle collaboration \cite{belle}, to observe the $0^+, 1^+$ states obtained from the $D^* \bar{K}^*$ interaction, where the $0^+$ state is associated to the $X_0(2866)$ state. From the eight reactions of the type $\bar{B} \to D^{(*)}  K^- K^{*0}$ of Ref. \cite{belle} we have selected two, the $\bar{B}^0 \to  K^{*0} D^{*+}  K^-$ and  $\bar{B}^0  \to K^0  D^+  K^-$, in order to observe the  $1^+$  and  $0^+$ states respectively. In the
first case the signal of the $1^+$ state stems from the original $\bar{B}^0 \to  K^{*0} D^{*+} K^{*-}$  reaction, followed by $D^{*+} K^{*-}$ interaction
to give the $R_1$ resonance,  which decays posteriorly to $D^{*+} K^{-}$. In the second case, the signal for the  $0^+$ state comes from the
$\bar{B}^0 \to  K^{0} D^{*+} K^{*-}$ with posterior $D^{*+} K^{-}$ interaction to produce the $0^+$ state, which decays lately into $D^+  K^-$. We could
relate the mass distributions of the signal and the background, finding very clear peaks for the $1^+$ and $0^+$ states. However,
we have argued that in the case of the $0^+$ state we expect the signal to be even more pronounced with respect to background than what is calculated here because of the suppressed $\bar{B}^0\to D^+K^-K^0$ decay versus $\bar{B}^0\to D^{*+}K^-K^{*0}$ decay at the tree level.

These reactions, already measured at Belle \cite{belle}, would need somewhat more statistics to show the  $D^{(*)+}  K^-$ peaks clearly. To facilitate the task
of determining the needed statistics, we have performed the integrated widths and we find
\begin{eqnarray}
&\Gamma(\bar{B}^0 \to  K^{*0} 1^+; 1^+ \to D^{*+} K^{-})/\Gamma(\bar{B}^0 \to  K^{*0} D^{*+} K^{-})=1.25\times 10^{-1}\,, \nonumber \\
&\Gamma(\bar{B}^0 \to  K^{0} 0^+; 0^+ \to D^{+} K^{-})/\Gamma(\bar{B}^0 \to  K^{0} D^{+} K^{-})=1.24\times 10^{-1}  \,.\nonumber
\end{eqnarray}

Together with the  branching ratios of the backgrounds of $12.9 \times 10^{-4}$, $1.6 \times 10^{-4}$ respectively for the former two
reactions \cite{belle},  the branching ratios for the exotic states produced stand well  within the experimental range presently accessible  in $B$ decays  \cite{pdg}. Thus, we can only encourage the performance of the experiment that should corroborate the existence of  the $X_0(2866)$  related to the existence of the
associated $1^+$ state. Since actually the $K \bar{K}^{*0}$ invariant  mass distribution  was already shown in Ref. \cite{belle} from accumulation of
events in different reactions, this means that one is not far from having enough  statistics to see the unmeasured  $D^{(*)+}  K^-$ peaks from
the individual reactions.  The other associated state $2^+$  is not easy to see with this kind of reactions and will have to wait its turn with different ones.

\section*{ACKNOWLEDGEMENT}
This work is partly supported by the National Natural Science Foundation of China under Grants Nos. 11975009, 12175066, 12147219.
RM acknowledges support from the program ``Contrataci\'on de
investigadores de Excelencia de la Generalitat valenciana'' (GVA)  with Ref. CIDEGENT/2019/015 and from the Spanish national grants PID2019-106080GB-C21 and PID2020-112777GB-I00.
This work is also partly supported by the Spanish Ministerio de Economia y Competitividad (MINECO) and European FEDER funds
under Contracts No. FIS2017-84038-C2-1-P B, PID2020-112777GB-I00, and by Generalitat Valenciana under contract PROMETEO/2020/023.
This project has received funding from the European Union Horizon 2020 research and innovation programme under
the program H2020-INFRAIA-2018-1, grant agreement No. 824093 of the STRONG-2020 project.

\end{document}